\documentstyle[eqsecnum,aps]{revtex}

\newcommand{\sig}{\tilde\sigma}
\newcommand{\ep}{\epsilon}
\newcommand{\ept}{\tilde\epsilon}
\newcommand{\be}{\begin{equation}}
\newcommand{\en}{\end{equation}}
\newcommand{\Do}{D^{(o)}}
\def\sutep{\rlap{\lower2ex\hbox{$\,\tilde{}$}}\epsilon{}}
\begin{document}
\draft
\title{Wave function of the Universe\\
and\\
Chern-Simons Perturbation Theory}
\author{Chopin Soo\cite{byline1}}
\address{Department of Physics,\\
National Cheng Kung University\\
Tainan 70101, Taiwan. }

\maketitle
\begin{abstract}
  The Chern-Simons exact solution of four-dimensional quantum gravity with
  nonvanishing cosmological constant is presented in metric variable
  as the partition function of a Chern-Simons theory with
  nontrivial source. The perturbative expansion is given,
  and the wave function is computed to the lowest order
  of approximation for the Cauchy surface which is topologically a
  3-sphere. The state is well-defined even at degenerate and vanishing
  values of the dreibein. Reality conditions for the Ashtekar variables
  are also taken into account; and remarkable features of the Chern-Simons
  state and their relevance to cosmology are pointed out.

\end{abstract}
 \widetext

\section{Overview}

   Soon after Ashtekar had recast and simplified the constraints of 4-d
gravity\cite{Ashtekar,Ashtekarbook,Ashtekarbook2}, it became
apparent the exponential of the Chern-Simons functional of the
connection variable is an {\it exact} solution\cite{Kodama} of
quantum gravity in the connection representation. Concurrent with
these developments, Witten in his seminal paper\cite{Witten}
showed how quantum field theory with non-Abelian Chern-Simons
action gave rise to 3-manifold and link invariants; and
subsequent work by many authors have made the Chern-Simons theory
one of the most successful topological theories to date. Much has
also been learnt from studying quantum gravity with the new
variables\cite{Ashtekarreview}.

  We begin with the observation that the Chern-Simons wave function is
  well-poised for the synthesis of these two strands of development;
because the state expressed in metric variable, is roughly(this
statement will be made precise in subsequent sections) the
partition function of the Chern-Simons action with the conjugate
variable as {\it source}.

  Although loop and spin-network transforms of the Chern-Simons state have been
  considered before\cite{Pullin-spin}, without assuming
  mini-superspaces as in Refs.\cite{Kodama,Paternogaetc}, the transformation to the
  more intuitive metric representation has been
  attempted, and computed for certain limits only fairly
  recently\cite{Paternoga}. Yet the
Chern-Simons state is arguably the most promising solution of
quantum gravity with the new variables discovered so far. Despite
being exact, it does not appear to suffer from the defects- among
them the lack of long range correlation and sensible continuum
limit\cite{Smolin-classical}- that affect many known
non-perturbative solutions. By virtue of being at the same time
the Hamilton-Jacobi functional\cite{Kodama,Smolin}, it should
lead to reasonable semi-classical correspondence. In fact, it is
also cosmologically interesting because it is associated with the
Cauchy data of constant curvature
3-manifolds\cite{Kodama,Kodamabook,Ezawa}(this assertion is
addressed briefly in Section IV). So even if the Chern-Simons
wave function is but one solution, it may be quite useful to
consider the simplifying assumption of the Chern-Simons state for
quantum gravity as a ``Quantum Cosmological Principle", analogous
to the classical assumption of ``The Cosmological Principle" with
its resultant Robertson-Walker metrics; and with regard to which
we may want to consider perturbations. This differs from just
doing mini-superspace quantum gravity because the Chern-Simons
wave function is an exact solution of the full theory. Most
remarkable of all its attributes is that when transformed to the
metric(more precisely, the densitized triad) representation, the
dimensionless coupling constant for the resultant Chern-Simons
theory is $\kappa=\frac{3}{8\pi\lambda{G}}$. Not only does this
offer an intriguing opportunity to entertain the cosmological
constant, $\lambda$, as a derived function of $\kappa$ and
Newton's constant, current astrophysical bounds place the value
of $\kappa$ at greater than $10^{120}$. This implies the usual
$\frac{1}{\sqrt\kappa}$ expansion\cite{Pullin-Gambini} for
Chern-Simons perturbation theory corresponds to a coupling of
roughly $10^{-60}$; so even perturbative results for the
Chern-Simons theory will be incredibly good, indeed more accurate
than for any other known physical theory. However, if we were to
entertain a Planck scale cosmological constant at, say, the
inflationary era or earlier, alternative strategies would be
required.

After briefly introducing Ashtekar's variables and establishing
the notations in Section II, the reality conditions are addressed
in Section III. One of the challenges to the synthesis of the
known results is how to reconcile the complex nature of Ashtekar's
variables for quantum gravity with ordinary real Chern-Simons
perturbation theory. Remarkably, the key to the resolution lies in
an inversion formula(to be discussed in Section III) discovered
many years ago\cite{Kodama,Kodamabook}, which says that in
transforming from the connection to the densitized triad
representation, the integration is not to be performed over all
complex values of the connection but only along a contour parallel
to the imaginary axis. It implies by a Wick rotation we can treat
the integration connection variable as real in the Chern-Simons
partition function. In Section IV  we recall the Chern-Simons
state and its semi-classical analog; and we adapt the known
results to write down the perturbative expansion for the
Chern-Simons wave function of quantum gravity in Section V. In
the final section, some relevant topological issues are covered,
including especially the effect of large gauge transformations,
and the chosen normalization; and we end by computing the lowest
order approximation of the wave function when the Cauchy manifold
is topologically a 3-sphere.

\section{Preliminaries}

Starting from real phase space variables $(\sig^{ia}, k_{jb})$
with the non-trivial Poisson bracket\footnote{We use natural
units $\hbar=c=1$, but retain $G$. Latin indices at the beginning
of the alphabet denote $SO(3)$ indices, while those from $i$
onwards are spatial indices.}
\be\{\sig^{ia}(\vec{x}),k_{jb}(\vec{y})\}_{P.B.} =
(8{\pi}G)\delta^i_j\delta^a_b\delta^3(\vec{x}-\vec{y}),\en
Ashtekar proposed the complex combination,
\begin{eqnarray} A_{ia}
&\equiv& ik_{ia} -\frac{1}{2}\ep_a\,^{bc}\omega_{ibc} \cr &=&
ik_{ia} + {{\delta F[\sig]} \over {\delta \sig^{ia}}} \label{1a}
\end{eqnarray}
and the densitized triad
$\sig^{ia}=\frac{1}{2}\epsilon^{abc}\ept^{ijk}e_{jb}e_{kc}$ as
the fundamental pair of variables for 4-d gravity. $\omega_{ab}$
is the spin connection compatible with the dreibein, $e_a$ i.e. $
de_a + \omega_{ab}\wedge e^b =0$; and modulo the constraints of
general relativity, $e^a_jk_{ia}$ is the extrinsic curvature.
Hence $A_{ia}$ transforms as a complex $SO(3)$ connection, and
its Poisson bracket with $\sig^{ia}$ is just
\be\{\sig^{ia}(\vec{x}),A_{jb}(\vec{y})\}_{P.B.} =
i(8{\pi}G)\delta^i_j\delta^a_b\delta^3(\vec{x}-\vec{y}).\en Since
$-\frac{1}{2}\ep_a\,^{bc}\omega_{ibc}$ is the functional
derivative of \be F[\sig]\equiv -\frac{1}{2}\int_M e^a \wedge
de_a, \en the Poisson bracket between two $A$ variables vanishes.
In other words, $F[\sig]$ is the sought-after generating
functional for the complex canonical
transformation\cite{Henneaux,Fukuyama,Kamimura} to
$(\sig^{ia},A_{jb}).$ Remarkably, the seven constraints for 4-d
gravity reduce to\cite{Ashtekar,Ashtekarbook,Ashtekarbook2} \be
D^{A}_i\sig^{ia} =0, \quad\quad \sutep_{ijk}\sig^{ia}{\tilde
B}^{j}_a, =0 {\label{2a}} \en and
\be\epsilon_{abc}\sutep_{ijk}\sig^{jb}\sig^{kc}\big({\tilde
B}^{ia} + \frac{\lambda}{3}\sig^{ia}\big)=0; {\label{2b}} \en
with $D^A$ denoting the covariant derivative with respect to
$A_{ia},{\tilde B}^{ia}\equiv\frac{1}{2}\ept^{ijk}F_{jka}$ the
magnetic field, and $\lambda$ the cosmological constant.

\section{Reality conditions, inner product, and an inversion formula}

The simplification of the constraints for 4-d space-times with
Lorentzian signature was achieved at the cost of complexification.
To recover the real phase space of general relativity, the
simplest procedure -but not the only one\cite{Ashtekarbook2}- is
to impose $\sig^\dagger_{ia} = \sig^{ia}$ and
$k^\dagger_{ia}=k_{ia}$ on the new phase space. The latter
reality condition on $k_{ia}$ translates into  \be
A^\dagger_{ia}= -A_{ia} +2{{\delta F[\sig]}
\over{\delta\sig^{ia}}}.\en Although somewhat complicated to
implement in the $A$-representation (the measure becomes
non-local then), it was pointed out in
Refs.\cite{Fukuyama,Kodama,Kodamabook}, and discussed in
Ref.\cite{Ashtekarbook2}, that the generating functional
$F[\sig]$ is known explicitly; so the reality condition on
$A_{ia}$ is straightforward to uphold in the
$\sig$-representation, and merely amounts to multiplying the
original measure by $\exp\big(\frac{-2F[\sigma]}{8{\pi}G}\big)$.
In other words, if we realize the original commutation relations
in the $\sig$-representation by \be\widehat{\sig}^{ia}\Psi[\sig]
=\sig^{ia}\Psi[\sig],\quad\quad
\hat{k}_{ia}\Psi[\sig]=(8{\pi}G){\delta\over{i\delta\sig^{ia}}}\Psi[\sig];\en
then
\be\hat{A}_{ia}\mapsto(8{\pi}G)\exp\big(-\frac{F[\sig]}{8{\pi}G}\big)
{\delta\over{\delta\sig^{ia}}}
\exp\big(\frac{F[\sig]}{8{\pi}G}\big);\en and it follows that the
action of $\hat{A}$ on
$\exp\big(\frac{F[\sigma]}{8{\pi}G}\big)\Psi[\sig]$,  is just
$\hat{A}_{ia}\mapsto (8{\pi}G){\delta\over{\delta\sig^{ia}}}$.
Moving over to the conjugate $A$-representation where
$\widehat{\sig}^{ia}\Psi[A]$ =
$-(8{\pi}G){\delta\over{{\delta}A_{ia}}}\Psi[A]$, the relation
between $\Psi[A]$ and
$\exp\big(\frac{F[\sigma]}{8{\pi}G}\big)\Psi[\sigma]$ is just the
usual Fourier transform\cite{Kodama,Kodamabook,Ashtekarbook2} \be
\Psi[A] =\int
D\sig\,\exp\big[\frac{1}{8{\pi}G}\big(F[\sig]-\int_M\sig^{ia}A_{ia}d^3x
\big)\big]\Psi[\sig].\label{3c}\en To express $\Psi[\sig]$ in
terms of $\Psi[A]$, it is important to observe
\cite{Kodama,Kodamabook,Ashtekarbook2} the inversion is {\it not}
by the naively integrating over {\it all} complex values of A,
but only over a contour, $C$, {\it parallel to the imaginary
axis}\footnote{The inversion prescription in
Refs.\cite{Kodama,Kodamabook} is for integration along the real
axis because the convention therein is $A=k+i\omega$. Our
$A=ik-\omega$ is similar to the original variable in
Ref.\cite{Ashtekar}, and corresponds to integration parallel to
the imaginary axis. If the integrand in Eq.(\ref{3a}) is
holomorphic in $A$, then the contour can be deformed. However, the
Chern-Simons functional is neither bounded from above nor below. }
\be\Psi[\sig] =\int_C DA\,\exp\big[\frac{1}{8{\pi}G}\big(-F[\sig]
+\int_M{\sig}^{ia}A_{ia}\big)\big]\Psi[A].\label{3a}\en This can
be seen by decomposing $A =iIm(A) + Re(A)$ in Eq.(\ref{3c}) to
give
\begin{eqnarray}
\int D[Im(A)]\,\exp\Big\{\frac{1}{8{\pi}G}\big[-F[\sig]
+i\int_M\sig{\cdot}Im(A)\big]\Big\}\Psi[A]&=& \int
D\sig'\,\Big\{\exp\big[\frac{1}{8{\pi}G}\big(F[\sig']-F[\sig]
-\int _M\sig'{\cdot}Re(A)\big)\big]\cdot\cr &&\int
D[Im(A)]\,\exp\big[i\int_M(\sig-\sig'){\cdot}Im(A)
\big]\Psi[\sig']\Big\}.
\end{eqnarray}
It is the integration over $Im(A)$ which gives
$\delta(\sig'-\sig)$, and the resultant inversion formula of
Eq.(\ref{3a}). Note that the inversion formula is not particular
to the Chern-Simons state, but applicable in general; and
suggests a strategy to circumvent the issue of the complex
connection variable by transforming to the conjugate
representation.

The factor $\exp\big(\frac{-F[\sigma]}{8{\pi}G}\big)$ has been
accounted for in Eq.(\ref{3a}), and the inner product for
$\Psi[\sig]$ is just\cite{Kodama,Kodamabook,Kamimura} \be \langle
\Psi|\Psi\rangle = \int D\sig\, \overline{\Psi[\sig]}\Psi[\sig].
\en Since \be\exp\big[\frac{1}{8{\pi}G}\big(-F[\sig]
+\int_M{\sig}^{ia}A_{ia}\big)\big] =
\exp\big[\frac{1}{16{\pi}G}(\int e^a\wedge D_Ae_a)\big], \en it is
explicitly diffeomorphism and gauge invariant\cite{Paternoga}.


\section{The Chern-Simons State}

  In the connection representation with $\hat{\sig}^{ia} \mapsto
  -(8{\pi}G){\delta\over{\delta{A}_{ia}}}$, we may express
   the operator
   \begin{eqnarray} \hat{Q}^{ia} &\equiv& {\tilde B}^{ia}
   + \frac{\lambda}{3}\hat{\sig}^{ia}\cr
   &=&-\exp\big(\kappa{CS[A]}\big){\delta\over{\kappa\delta{A_{ia}}}}
\exp\big(-\kappa{CS[A]}\big),
\end{eqnarray}
in terms of the Chern-Simons functional
 \be CS[A] = \frac{1}{2}\int_M (A^a \wedge dA_a +
\frac{1}{3}\ep^{abc}A_a\wedge A_b\wedge A_c) \en since
$\frac{{\delta}CS[A]}{\delta{A_{ia}}}={\tilde B}^{ia}$ if
$\partial M =0$. Here $\kappa \equiv \frac{3}{8{\pi}\lambda{G}}$
is the dimensionless parameter mentioned previously. For
non-vanishing cosmological constant, the constraints of
Eqs.(\ref{2a}) and (\ref{2b}) can be rewritten as \be
D^{A}_i\hat{Q}^{ia}\Psi[A] =0,
 \quad\quad \sutep_{ijk}{\hat{\sig}}^{j}_a\hat{Q}^{ia}\Psi[A]
=0\en and
\be\epsilon_{abc}\sutep_{ijk}{\hat{\sig}}^{jb}{\hat{\sig}}^{kc}
\hat{Q}^{ia}\Psi[A]=0.\en It follows, for this ordering, that a
sufficient condition for an {\it exact} state $\Psi[A]$ to
satisfy all the constraints of 4-d quantum gravity is that it is
annihilated by $\hat{Q}^{ia}$\cite{soo,Chang}. It is easy to
check the solution is \be \Psi[A]={\cal
N}\exp\big(\kappa{CS[A]}\big), \quad\quad \frac{\delta{\cal
N}}{{\delta}A_{ia}}=0;\label{3d} \en with ${\cal N}$ being a
topological invariant, as indicated in Eq.(\ref{3d}). We shall
refer to this state as the ``Chern-Simons state". In Section VI
we shall argue that $\cal N$ is needed to provide a compensating
topological factor under large gauge transformations. Since we are
dealing with quantum gravity, we cannot rule out summing over
topologically inequivalent closed 3-manifolds and bear in mind
the general solution \be \Psi[A]=\sum_{Top (M):{\partial{M}=0}
}{\cal N}_M\exp\big(\kappa{CS[A]}_M\big).\en

 It is most remarkable that $G$ and $\lambda$ has come together
 as the dimensionless coupling constant $\kappa$.
 Thus in the Chern-Simons perturbation
theory to be discussed the cosmological constant may be
considered as a derived quantity from $G$ and $\kappa$. Note
however that without further restrictions, the Chern-Simons
functional is invariant under arbitrary small gauge
transformations only if the 3-dimensional Cauchy surface $M$ is
without boundary ($\partial M=0$); otherwise its functional
derivative contains boundary effects, and the exponential of the
Chern-Simons functional fails to solve the constraints.

We note in passing that in the connection representation,
$\lambda=0$ is a singular limit, since it changes the functional
differential equation of the super-Hamiltonian constraint of
Eq.(\ref{2b}) from third order to second.
 With regard to convergence, the perturbation series to be
discussed is not any less well-defined for one sign of $\kappa$
as for the other. This is also true for the coupling constant of
a normal Chern-Simons theory. However, the theory at hand refers
to gravity, and the coupling or cosmological constant is, at
least semi-classically, not entirely independent of the topology
of M. We note that $CS[A]$ is at the same time a Hamilton
function for the classical Hamilton-Jacobi theory, since replacing
$\sig^{ia}$ by $-(8{\pi}G)\frac{\delta}{\delta
A_{ia}}(\kappa{CS[A]}) =-\frac{3}{\lambda}{\tilde B}^{ia}$ solves
all the classical constraints. This Cauchy data is also known as
the Ashtekar-Renteln ansatz\cite{Renteln}. Under evolution, the
ansatz describes classical Einstein manifolds with vanishing Weyl
two-form i.e. $W_{AB} = R_{AB}- {\lambda \over 3}e_A\wedge e_B
=0$; with $e_A$ denoting the vierbein in four dimensions, and
$R_{AB}$ the Riemann curvature two-form. Solutions with positive
$\lambda$ include the de Sitter manifold. On 3-d Cauchy surfaces,
 the Ashtekar-Renteln ansatz which
should correspond to the semi-classical limit of the state,
implies (with $A$ as in Eq.(\ref{1a})) \be iD_\omega k_a=0,
\quad\quad R_{ab}+ k_a\wedge k_b = \frac{\lambda}{3}e_a\wedge
e_b, \en on equating the real and imaginary parts. Solutions of
the first equation include $k_a =\alpha e_a$ with constant
$\alpha$, which implies M is a 3-manifold with {\it constant}
Riemannian curvature $R_{ab}=(\frac{\lambda}{3} -\alpha^2)
e_a\wedge e_b$ and Ricci scalar $R =2(\lambda -3\alpha^2)$. Since
two constant curvature surfaces with the same value of $R$ are
isometric\cite{Eisenhart}, the simply-connected constant
curvature 3-manifolds are exhausted by $S^3, R^3$ and $H^3$ with,
respectively, $+,0$ and $-$ curvature. Thus to the extent
$\partial M=0$ is required, the Chern-Simons state not only
selects $S^3$ as the only closed alternative, but also implies
the cosmological constant $\lambda$ is positive since $R > 0$ for
$\partial M =0$; in which case $\alpha^2 < \frac{\lambda}{3}.$
The actual 3-topology of our universe is not yet settled
\cite{Luminet}; but to the extent that the Chern-Simons state is
capable of describing our universe by association with the de
Sitter phase\cite{Kodama,Smolin} of inflationary
expansion\cite{Guth}, of the three alternatives for
simply-connected Robertson-Walker space-times, it would, barring
topology changes during evolution after the inflationary phase, in
fact be compatible only with the {\it closed} model for the
3-topology of our present universe.

\section{The wave function as a Chern-Simons perturbation series}

  We shall now employ functional methods to expand the Chern-Simons state
  in the $\sig$-representation as a Chern-Simons perturbation series.
Details on Chern-Simons perturbation theory can be found in
Refs.\cite{Axelrod,Bar-Nataan,Rozansky}; and fortuitous factors
allow the adaptation of these results to the case at hand, even
though the connection variable for 4-d quantum gravity is complex.

   From Eqs.(\ref{3a}) and (\ref{3d}), the Chern-Simons state in the
$\sig$-representation is \be \Psi[\sig] = {\cal N}\int_{C} DA \,
\exp\{(\frac{1}{16{\pi}G}\int_M e^a \wedge D_Ae_a) + \kappa CS[A]
\}. \label{3e} \en We leave $\cal N$ unspecified for the moment,
and take the liberty of absorbing integration constants into it
when the need arises. It will be brought up more concretely later
on.

 Writing the connection as\footnote{If $\kappa$ is negative, we
 use ${A^{(o)}} +\frac{1}{\sqrt{|\kappa|}}a$ instead.} $ A = {A^{(o)}}+
\frac{1}{\sqrt{\kappa}}a$ leads to \be \Psi[\sig] = {\cal N}
\exp\big\{\big(\frac{1}{16{\pi}G}\int_M e^a\wedge{\Do}e_a\big) +
\kappa CS[A^{(o)}]\big\}Z[\sig, A^{(o)}], \label{4bb}\en with \be
Z[\sig, A^{(o)}]=
 \int_C Da \,
\exp\{\frac{1}{2}\int_M [a^a\wedge{\Do}a_a +
{2\sqrt{\kappa}}(F^a_{A^{(o)}} +
\frac{\lambda}{6}\ep^{abc}e_b\wedge e_c)\wedge a_a
+\frac{\ep^{abc}}{3\sqrt{\kappa}}a_a\wedge a_b\wedge a_c]\}.
\label{4b} \en This is true for {\it any} $A^{(o)}$. The question
is how best to compute $Z[\sig, A^{(o)}]$. It is customary to
choose a stationary point to eliminate the linear term in $a$;
for the case at hand, this means $A^{(o)}$ satisfies
$F^a_{A^{(o)}} + \frac{\lambda}{6}\ep^{abc}e_b\wedge e_c=0$.
However this leads to complications. Without the luxury of
constraining $e_a$, solving for $A^{(o)}$ in terms of the
dreibein implies a non-Hermitian quadratic term which is
furthermore a complicated function of $\sig$. Moreover, a linear
term in $a$ will anyway appear as a ghost-antighost-gauge
connection coupling after gauge fixing.

Happily, the expression for $\Psi[\sig]$ also suggests we treat
$\sig$ as the {\it source} for $A$; and expand about a {\it flat
connection} $F^a_{A^{(o)}}=0 $ i.e. $A^{(o)}= UdU^{-1}$ locally.
It follows that \be Z[\sig, A^{(o)}]=\int_C Da \,
\exp\{\frac{1}{2}\int_M [a^a\wedge{\Do}a_a
+\frac{\ep^{abc}}{3\sqrt{\kappa}}a_a\wedge a_b\wedge
a_c]+\frac{1}{8{\pi}G\sqrt{\kappa}}\int_M\sig^{ia} a_{ia} d^3x
\}. \en

The Chern-Simons functional is invariant under gauge and general
coordinate transformations. To compute its generating functional
through Gaussian integrals, gauge-fixing is required. This
entails the introduction of an auxiliary metric $g^{ij}$ which is
{\it independent of}, and not to be confused with, the source
$\sig^{ia}$. If the gauge-fixing action is a
Becchi-Rouet-Stora-Tyutin(BRST) variation, nilpotency of the BRST
operation guarantees that the full action is BRST-invariant. To
compute the generating functional, it is sufficient to take care
of just the Yang-Mills invariance by adding the gauge fixing
action
\begin{eqnarray} \int_M L_{gauge-fixing} &=&\int_M \delta_{BRST}
[{\bar{c}}^a({\Do}*a_a -*\xi b_a)]\cr &=& -\int_M
[\frac{1}{\sqrt{\kappa}}(b^a{\Do}*a_a +
{\bar{c}}^a({\Do}*[a,c])_a)+{\bar{c}}^a{\Do}*{\Do}c_a +*\xi
b^ab_a].
\end{eqnarray}
$*{\Do}*{\Do}c_a =\triangle c_a$ is just the action of the
Laplacian twisted by the flat connection, $\triangle$, acting on
the ghost; and $*$ is the Hodge dual operator which depends on
$g^{ij}$.

After gauge-fixing, the generating functional,
\begin{eqnarray} Z[\sig, A^{(o)}]=\int_C da \int d\bar{c} \int dc\,
&&\exp\{\frac{1}{2}\int_M (a^a\wedge D_{A_{(o)}}a_a
+\frac{\ep^{abc}}{3\sqrt{\kappa}}a_a\wedge a_b\wedge
a_c-\frac{1}{\sqrt{\kappa}}{\bar{c}}^a({\Do}*[a,c])_a-*{\bar{c}}^a\triangle
c_a) \cr &+&\frac{1}{8{\pi}G{\sqrt\kappa}}\int_M\sig^{ia} a_{ia}
d^3x\} \int db\, \exp\{\frac{-1}{\sqrt{\kappa}}\int_M
(b^a{\Do}*a_a+*\xi b^ab_a)\},
\end{eqnarray}
includes integrations over $(c,\bar{c})$ ghost-antighost; and the
auxiliary field $b$. The Gaussian integral over $b$ is just
 \be
 \int
db\, \exp\{\frac{-1}{\sqrt{\kappa}}\int_M (b^a{\Do}*a_a+*\xi
b^ab_a)\} =
C_\xi\exp\{-\frac{1}{4\kappa\xi}\int_M(*a^a)\wedge{\Do}*{\Do}*a_a
\} \en after integration by parts.

If we were forced to integrate over all complex values of $a$, the
generating functional would not be well-defined. Fortunately, as
was discussed in an earlier section, the integration contour $C$
is only parallel to the imaginary axis. For convenience we choose
the constant real part of the contour to be just $A^{(o)}$. As a
result, the integration over $A$ along $C$ can be converted into
integration over $a$ along the {\it real} axis $R$ by a simple
Wick rotation, $a \mapsto -i{a}$. Hence the generating functional
assumes the form
\begin{eqnarray} Z[\sig, A^{(o)}]=\int_R da \int d\bar{c} \int dc\,
&&\exp\{-\frac{1}{2}\int_M [a^a\wedge ({\Do}
-\frac{1}{2\kappa\xi}*{\Do}*{\Do}*)a_a
+\frac{i\ep^{abc}}{3\sqrt{\kappa}}a_a\wedge a_b\wedge
a_c\cr&-&\frac{i}{\sqrt{\kappa}}{\bar{c}}^a({\Do}*[a,c])_a+*{\bar{c}}^a\triangle
c_a] -\frac{i}{8{\pi}G{\sqrt\kappa}}\int_M\sig^{ia} a_{ia} d^3x
\}.
\end{eqnarray}
Apart from the difference of a factor of $i$ in the quadratic term
in $a$, the expression is identical to the gauge-fixed generating
functional of a Chern-Simons theory for a {\it real} $SU(2)$
connection. $J^{ia} \equiv
\frac{1}{8{\pi}G{\sqrt\kappa}}\sig^{ia}$ is manifestly the source
for $a_{ia}$.

An upshot of gauge-fixing is that operator $L$ in the quadratic
term in $a$,
 \be \int\int dxdy\, a_{ia}(x) L^{ikab}(x,y) a_{kb}(y)
 \equiv -\frac{1}{2}\int\int dxdy\,
a_{ia}(x)\delta(x-y)(\ept^{ijk}{\Do}_j-\frac{1}{2\kappa\xi}{\Do}_l
g^{il} {\Do}_m\sqrt{g}g^{km})^{ab}a_{kb}(y),\label{se} \en is
invertible\footnote{Zero modes of $\triangle$ and $L$ are taken
into account carefully in Ref.\cite{Rozansky}; and the effect of
reducible connections on the Faddeev-Popov determinant is
discussed in \cite{Adams,AdamsRev}. Here we assume the zero modes
have been subtracted, and the operators are invertible.}. The
expedient choice of flat metric for $g^{ij}$ yields the inverse
as \be
L^{-1}_{ijab}(x,y)=\frac{2}{\triangle}\{\epsilon_{ijk}{\Do}^k
-\frac{2\kappa\xi}{\triangle}{\Do}_i{\Do}_j\}_{ab}\delta(x,y)
\label{5d} \en

The cubic term in $a$ and the ghost-antighost-gauge interaction
both carry a $\frac{1}{\sqrt\kappa}$ factor, and will therefore be
addressed perturbatively by expanding in powers of
$\frac{1}{\sqrt\kappa}$. To wit, we introduce sources $\bar\eta$
and ${\eta}$ for $c$ and $\bar{c}$ respectively; and express the
generating functional as
\begin{eqnarray}
Z[J, A^{(o)}]&=&\Big[\exp\big(\frac{1}{\sqrt{k}}\int_M
\ep^{abc}\big[\frac{i}{2}({\Do}^i{\partial \over {\partial
\eta}})_a{\partial \over {\partial J^{ib}}}{\partial \over
{\partial {\bar{\eta}}^c}} + \frac{\ept^{ijk}}{3}{\partial \over
{\partial J^{ia}}}{\partial \over {\partial J^{jb}}}{\partial
\over {\partial J^{kc}}}\big]\big)\cr &&\int d\bar{c} \int dc
\exp[\int_M *(-{\bar{c}}^a\frac{\triangle}{2}c_a +{\bar\eta}^ac_a
+ \bar{c}^a\eta_a)]\cdot\cr &&\left.\int_R da
\exp\big([-\frac{1}{2} \int_M\int_M dxdy\,a_{ia}(x)
L^{ijab}(x,y)a_{jb}(y)] -i\int_M J^{ia} a_{ia}
d^3x\big)\Big]\right|_{\bar{\eta}=\eta=0}.
\end{eqnarray}
Again, the Gaussian integrals over ghost-antighost variables, and
over real $a$, can be performed readily, leading to
\begin{eqnarray} Z[J, A^{(o)}]&=& \frac{\det(\triangle)}{\sqrt{\det(L)}}
\Big[\exp\big(\frac{1}{\sqrt{k}} \int_M
\ep^{abc}\big[\frac{i}{2}({\Do}^i{\partial \over {\partial
\eta}})_a{\partial \over {\partial J^{ib}}}{\partial \over
{\partial {\bar{\eta}}^c}} + \frac{\ept^{ijk}}{3}{\partial \over
{\partial J^{ia}}}{\partial \over {\partial J^{jb}}}{\partial
\over {\partial J^{kc}}}\big]\big)\cdot\cr && \left.\exp(\int_M
*{\bar{\eta}}^a\frac{2}{\triangle}\eta_a)
\exp\big(-\frac{1}{4}\int_M dx\int_M dy\,J^{ia}(x)
L^{-1}_{ijab}(x,y)J^{jb}(y)\big)\Big]\right|_{\bar{\eta}=\eta=0}.
\label{4a}
\end{eqnarray}
Note that the expression is defined for degenerate as well as
non-degenerate dreibeins.

We shall be interested in computing the ratio
$Z[J,A^{(o)}]/Z[J=0,A^{(0)}]$ rather the absolute value of the
generating functional. It is also permissible to assume the
``Landau gauge" $\xi=0$ after the gauge fixing. This corresponds
to imposing the condition ${\Do}*a =0$; and has the geometrical
interpretation physical excitations are orthogonal (with respect
to the metric $g^{ij}$) to gauge variations i.e. $\int_M (* a)^a
\wedge(\delta_{BRST} A)^{(o)}_a =0.$ It also has the advantages of
not introducing the external spurious scale $\xi$ and avoiding
infrared divergences \cite{Alvarez-Gaume,Delduc,Blasi}.

Naively, in the Chern-Simons generating functional the auxiliary
metric enters only through gauge fixing, so the stress tensor,
which captures the dependence on $g^{ij}$, is formally a BRST
commutator. However proof of full metric independence can be very
involved. Ref.\cite{Bar-Nataan-Witten} discusses the metric
dependence when regularization effects are also taken into
account, and Ref.\cite{Axelrod} contains a formal proof of metric
independence. In this article, we do not investigate the metric
dependence or independence beyond noting that the additional
ingredient here is the source term which requires no reference to
$g^{ij}$ since $\sig^{ia}$ is a densitized vector. Thus the metric
dependence of the results is arguably no worse than for the usual
Chern-Simons theory with source. This topic certainly merits more
careful study, especially when the aim here is to reconstruct
quantum geometrodynamics from quantum gauge dynamics. The present
article should be regarded as a first attempt at systematically
computing the Chern-Simons state in the metric representation via
Chern-Simons perturbation theory.

\section{Discussions and further remarks}

The limit for zero source is taken to be $\left.Z[J=0,A^{(o)}]=
Z[J, A^{(o)}]\right|_{J=0}$. As we shall see, for {\it vanishing}
dreibein ($e_a =0$) the wave function $\Psi[0] = {\cal
N}\exp(\kappa CS[A^{(o)}])Z[J=0, A^{(o)}]$ may be well-defined,
even if classical 3d-manifolds with degenerate dreibeins were
problematic. It is not entirely clear what ``normalization
factor" ($\cal N$ of Eq.(\ref{3e})) should be adopted; but we
shall take the following considerations into account.

Unlike the {\it partition function}, $\int
DA\,\exp(i{\kappa}CS[A])$, of the usual Chern-Simons theory, {\it
no} quantization condition on $\kappa$ for the {\it wave function}
of Eq.(\ref{3d}) (or Eq.(\ref{3e})) arises\footnote{This does not
imply there is no renormalization of $\kappa$. In normal
Chern-Simons theory, quantum corrections results in a shift of
the coupling constant. However, if we are computing the ratio
$Z[J,A^{(o)}]/Z[J=0,A^{(o)}]$, any counter term, e.g.
$\exp\big(\kappa'CS[A^{(o)}]\big)$, should cancel out if it is
identical for the theory with and without source.}. Strictly
speaking, a {\it state} needs only to be invariant up to a phase
under large gauge transformations - an example is the
$\theta$-vacuum\cite{Jackiw}. Moreover, there is a difference of a
factor of i; so for the Chern-Simons state here large gauge
transformations, which induces $CS[A] \mapsto CS[A] + 8\pi^2n$,
multiplies the wave function by $\exp(8\pi^2n\kappa)$ instead a of
phase factor. Without a compensating topological factor in $\cal
N$, the naive Chern-Simons solution is therefore not even
invariant up to a phase under large gauge transformations.
Furthermore, since its magnitude can be made arbitrarily big by
translation of the connection under large gauge transformations,
the Chern-Simons functional is neither bounded from above nor
below. These complications can however be neutralized by
accommodating $\exp(\kappa CS[A^{(o)}])$ in the denominator of
$\cal N$ to cancel out the effect of large gauge
transformations\footnote{Ref.\cite{Paternoga} uses a similar
normalization, by dividing with the exponent of the winding
number. By expanding about a flat connection, the topological
invariant $\exp(\kappa CS[A^{(o)}])$ arises naturally in our
present discussion.}. This means that we are in effect considering
the state to be proportional to $\exp[{1\over {(16\pi G)}}\int
e^a \wedge D^{(o)}e_a]Z[\sig, A^{(o)}]$. Since $a$ transforms
covariantly, $Z[\sig, A^{(o)}]$ of Eq.(5.3) clearly implies the
resultant state will be invariant under {\it both} small {\it
and} large gauge transformations. Note also that although $CS[A]$
is not a topological invariant for arbitrary connections, the
Chern-Simons functional of a flat connection is. In quantum field
theories, $\cal N$ usually carries the normalization factor
 $(Z[J=0])^{-1}$ (normalized generating functionals
 eliminate vacuum diagrams). For the case at hand we can
 opt to include $\exp(\kappa CS[A^{(o)}])Z[J=0, A^{(o)}]$ in the
 denominator of $\cal N$.

The absolute value of the ratio of the determinants in
Eq.(\ref{4a}) is the square root of the Reidemeister-Ray-Singer
analytic torsion, $\tau_{(o)}$, of the flat connection
$A^{(o)}$\cite{Schwartz}. Since the Laplacian operator is
positive-definite, the ratio, including the phase which comes
only from $L$, is expressible as \be
\frac{\det(\triangle)}{\sqrt{\det(L)}} =
\tau^{\frac{1}{2}}_{(o)}\exp\big(-i\pi\eta_L[A^{(o)}]\big), \en in
which the spectral asymmetry of $L$ is accounted for by a suitably
regularized eta-invariant\cite{Witten}. This ratio can be made
independent of the background metric by adding a phase, which is
the linear combination of the gravitational Chern-Simons
functional and the eta-invariant of $L$ coupled to $g^{ij}$ and
the trivial connection\cite{Witten}. However, if we are
interested in $Z[J,A^{(o)}]/Z[J=0, A^{(o)}]$, the ratio of the
determinants, together with its framing dependence on $g^{ij}$,
{\it cancels out} since it appears in both the numerator and
denominator.\footnote{If $M$ supports more than one distinct flat
connection, we should sum, or integrate, over each contribution in
Eq.(\ref{4bb}) and may want to consider ${\cal N}^{-1}= \sum_{(o)}
\exp(\kappa CS[A^{(o)}])Z[J=0,A^{(o)}]$.}

Taking the above considerations into account, and for
concreteness, we choose $\cal N$ so that  the wave function \be
\Psi[J, A^{(o)}]= \exp\big(\frac{1}{16{\pi}G}\int\,
e^a\wedge{\Do}e_a\big){{Z[J, A^{(o)}]}\over {Z[J=0,A^{(o)}]}},
\en with $J^{ia}\equiv
\frac{1}{16{\pi}G{\sqrt\kappa}}\epsilon^{abc}\ept^{ijk}e_{jb}e_{kc}$,
is invariant under both small and large gauge
transformations.\footnote{There remains of course the freedom to
multiply $\Psi[J,A^{(o)}]$ by a true topological invariant which
could come in useful when we consider its normalization for the
integration over $\sig$ in the norm.} This implies at vanishing
dreibein $\Psi[0]=1$.

  As an illustration, consider $M=S^3$ topologically; $A^{(o)}$
  is hence gauge equivalent to the trivial connection. Adopting a
  flat auxiliary metric and the Landau gauge,
the resultant generating functional is
\begin{eqnarray} Z[J, A^{(o)}]&&=\frac{\det(\triangle)}{\sqrt{\det(L)}}
\Big[\exp\big(\frac{1}{\sqrt{k}} \int
\ep^{abc}\big[\frac{i}{2}({\partial}^i{\partial \over {\partial
\eta}})_a{\partial \over {\partial J^{ib}}}{\partial \over
{\partial {\bar{\eta}}^c}} + \frac{\ept^{ijk}}{3}{\partial \over
{\partial J^{ia}}}{\partial \over {\partial J^{jb}}}{\partial
\over {\partial J^{kc}}}\big]\big)\cdot\cr &&
\left.\exp\big\{\int_M
{\bar{\eta}}^a\frac{2}{\triangle}\eta_a\big\}
\exp\big\{-\frac{1}{2}\int_M
\,\sutep_{ijk}J^{ia}\frac{{\Do}^j}{\triangle}J^{k}_a\big\}
\Big]\right|_{\bar{\eta}=\eta=0}. \label{4d}
\end{eqnarray}
The propagators for ghost-antighost, and for $a$; and the 3-point
vertices can be read off easily. On recalling
  the simple identities
\be \triangle (\frac{1}{|x-y|})=-4\pi\delta{(x-y)}, \quad\quad
  \partial^i\frac{1}{|x-y|}=-\frac{(x-y)^i}{|x-y|^3}\,,\en
 the two-point
function\footnote{This ``propagator" for the Chern-Simons theory
should not be confused with the expectation value,
$\langle{\Psi}|{\hat A}_{ia}(x){\hat A}_{jb}(y)|\Psi\rangle=
-\frac{1}{\kappa}\int d\sig\,
\overline{\big[\frac{\delta\Psi[\sig] }{\delta
J^{ia}(x)}\big]}\frac{\delta\Psi[\sig]}{\delta{J^{jb}}(y)} $ in
quantum gravity. Rather, the Chern-Simons ``propagator"  is a
computational device for the latter.}
 is\cite{Alvarez}
 \be \langle a_{ia}(x)a_{jb}(y)\rangle_J
=-\delta_{ab}\epsilon_{ijk}\frac{(x-y)^k}{8{\pi}|x-y|^3}.
\label{5p} \en
 Standard Feynman diagram techniques in quantum field theory can be applied
to compute the perturbative expansion, in
$\frac{1}{\sqrt{\kappa}}$, of the exponential operator with
functional derivatives in Eq.(\ref{4d}). To lowest order, we find
 \be \frac{Z[\sig,A^{(o)}]}{Z[\sig=0,A^{(o)}]}\approx
\exp\big\{\frac{-1}{2^9{\kappa}G^2\pi^3}\int dx \int dy
\,\epsilon_{ijk}\sig^{ia}(x)
\frac{(x-y)^j}{|x-y|^3}\sig^{ka}(y)\big\}.\label{5s} \en
exhibiting inverse square law long range correlation. Note also
that $\lim_{x\rightarrow
y}\epsilon_{ijk}\sig^{ia}(x)\sig^{ka}(y)=0$ curbs the singular
behaviour from coincident points. However, we observe that
Eq.(\ref{5s}) is not explicitly gauge invariant; and the often
cited propagator of Eq.(\ref{5p}) does not transform in an
explicitly covariant manner. Following a suggestion of Schwinger
many years ago\cite{Schwinger}, the situation can be rectified
with a phase factor, ${P}\big(\exp \int^x_y A^{(o)}\big)$, along
a path connecting $x$ and $y$. It is to be noted this factor here
is {\it independent} of the path, no matter how far separated the
points $x$ and $y$ are, because $A^{(o)}$ is a {\it flat}
connection and every closed loop in $S^3$ is contractible. The
justification is that careful computations with $A^{(o)}=
UdU^{-1}$ in Eqs.(\ref{se}) and (\ref{4d}), indeed produces an
extra factor of $U(x)U^{-1}(y) = P(\exp\int^y_x A^{(o)})$ which
takes the place of the $\delta_{ab}$ in the propagator in
Eq.(\ref{5p}).

Since $A^{(o)}$ is flat, the factor $\int_M e^a\wedge{\Do}e_a$
can also be written in terms of Chern-Simons functionals and the
volume of $M$ as \be \frac{1}{2l^2_p} \int_M e^a\wedge{\Do}e_a =
CS[A^{(o)}+(e/{l_p})]-CS[A^{(o)}]-\frac{1}{6l^3_p}\int_M
\ep^{abc}e_a\wedge e_b\wedge e_c,\en with the first term denoting
the Chern-Simons functional of $A^{(o)}_a$ shifted by $e_a$
divided by the Planck length $l_p=\sqrt{G}.$ Combining these
previous expressions, to the lowest order,
\begin{eqnarray}
\Psi[\sig]\approx&& \exp\big\{\frac{1}{8\pi}\big(
CS[A^{(o)}+({e}/{l_p})]-CS[A^{(o)}]-\frac{V_M}{l^3_p}\big)\big\}\cdot\cr
&&\exp\big\{\frac{-1}{2^9{\kappa}l^4_p\pi^3}\int dx \int dy
\,\epsilon_{ijk}\sig^{ia}(x)[P(\exp{\int^x_y
A^{(o)}})]_{ab}\sig^{kb}(y)\frac{(x-y)^j}{|x-y|^3}\big\}.
\label{5a} \end{eqnarray} There is still a long way to go before
the quantum fluctuations contained in Eq.(\ref{5a}) can be tested,
say, in terms of the cosmic microwave radiation. At the very
least we do not expect a finite norm for $\Psi[\sig]$. The
non-normalizability of the Chern-Simons state has been
demonstrated explicitly in a mini-superspace model\cite{Marugan},
and in a certain limit after gauge fixing\cite{Paternoga}. In
fact, the divergence of the norm is to be expected on general
grounds\cite{Time}. Some have argued until we confront ``the
problem of time"\cite{Conceptual}, a probability amplitude
interpretation for $\Psi[\sig]$ does not make sense; since
integrating over $\sig$ amounts to summing over all physical
``times". Thus naive ``non-normalizability" of a solution of the
constraints does not imply spuriousness. Deciding what a wave
function of the universe really means forces us to confront deep
conceptual and interpretive issues\cite{Conceptual}. Being exact
despite its simplicity, computable systematically in familiar
metric terms, semi-classically relevant and cosmologically
interesting; the Chern-Simons wave function can be a valuable
proving ground for these quantum gravity issues and their
proposed resolutions.

In the ``slow-roll" stage of inflationary scenarios\cite{peacock},
the effective cosmological constant of the de Sitter expansion
phase can be estimated by ${\lambda_{eff.}\over {(8\pi G)}} \sim
\sigma_{S.B.}T^4_{GUT}$, where $\sigma_{S.B.}$ denotes the
Stefan-Boltzmann constant. This corresponds to an effective
Chern-Simons coupling of $\frac{1}{\sqrt{\kappa_{eff.}}} \sim
10^{-7}-10^{-5}$ if the grand unification energy scale is taken to
be $10^{15}-10^{16}$ GeV. So Chern-Simons perturbation theory
should be applicable during this stage. Moreover, since this grand
unification scale is several orders below the Planck regime,
semi-classical gravity is expected to dominate although it may
still be necessary and desirable to include quantum gravity
fluctuations. Thus the perturbative Chern-Simons wave function
discussed here may be most pertinent to the inflationary period
of our universe.

\acknowledgments I would like to thank Louis Kauffman and Lee
Smolin for helpful discussions and correspondence during my work
on the Chern-Simons wave function. The research for this work has
been supported in part by funds from the National Science Council
of Taiwan under Grant Nos. NSC 89-2112-M-0060-050 and
90-2112-M-006-012.


\end{document}